%% file: proceedings_arxiv.tex
\documentclass{aip-cp}

\usepackage[numbers,sort&compress]{natbib}
\usepackage{rotating}
\usepackage{graphicx}
\usepackage{color}
\usepackage{xspace}
\usepackage{units}
\makeatletter %
\let\iint\@undefined
\let\iiint\@undefined
\let\iiiint\@undefined
\let\idotsint\@undefined
\makeatother
\usepackage{amsmath} %
\usepackage{url} %

\input{befehle_arxiv}

\graphicspath{{figures/}}

\newcommand{\com}{center-of-mass\xspace}
\newcommand{\incgraphics}[2]{\includegraphics[#1]{#2}}
\newcommand{\Fig}   [1]{Fig.~\ref{fig:#1}} %

\newcommand{\supersymmetry}{supersymmetry\xspace} %

\newcommand{\ConeNtwo  }{\ensuremath{\charginon{1}\neutralinon{2}}\xspace}
\newcommand{\ConeOS    }{\ensuremath{\charginon{1}\charginomp_1}\xspace}
\newcommand{\ConeSS    }{\ensuremath{\charginon{1}\charginon{1}}\xspace}
\newcommand{\NtwoNthree}{\ensuremath{\neutralinon{2}\neutralinon{3}}\xspace}

\newcommand{\analysis}[1]{\textit{#1}} %

\begin{document}
\title{Search for ``Electroweakinos'' \\with the ATLAS Detector at the LHC}

\author[aff1]{Alexander Mann\corref{cor1}, on behalf of the ATLAS Collaboration}

\affil[aff1]{Ludwig-Maximilians-Universit\"at M\"unchen, Fakult\"at f\"ur Physik, Am Coulumbwall 1, 85748 Garching bei M\"unchen.}
\corresp[cor1]{Corresponding author: A.MannLMU.de}

\maketitle

\begin{abstract}
  Supersymmetry is one of the most popular extensions of the Standard Model of particle physics, 
  as it offers solutions to several shortcomings of the Standard Model.
  Natural supersymmetric models favor masses for the new particles which are predicted by \supersymmetry in the range of hundreds of GeV, 
  well within the reach of the Large Hadron Collider at CERN. 
  If squarks and gluinos are much heavier, the production of charginos and neutralinos may be the dominant production mode for supersymmetric particles. 
  These proceedings present
  results from new searches for the production of charginos and neutralinos, 
  focusing %
  on the recent paper by the ATLAS collaboration %
  that summarizes and extends the searches for the electroweak production of supersymmetric particles using data from Run-1 of the LHC. %
\end{abstract}

\section{INTRODUCTION}
Supersymmetry (SUSY) 
  \cite{Miyazawa:1966,Ramond:1971gb,Golfand:1971iw,Neveu:1971rx,Neveu:1971iv,Gervais:1971ji,Volkov:1973ix,Wess:1973kz,Wess:1974tw} %
  is one of the most popular extensions of the Standard Model of particle physics, 
as it can provide solutions to a number of problems or short-comings of the Standard Model.
It introduces a new space-time symmetry between fermions and bosons
  and predicts essentially a doubling of the number of elementary particles contained in the model. 
A large number of searches have been designed and carried out in the past to find traces of these particles in collider experiments. 
These proceedings discuss the search for electroweakinos with the ATLAS detector~\cite{PERF-2007-01},  
  one of the two large multi-purpose detectors at the Large Hadron Collider (LHC)~\cite{Evans:2008zzb} %
  at CERN.
Electroweakinos comprise the supersymmetric charginos \charginon{i} ($i=1,2$) and neutralinos \neutralinon{j} ($j=1,\dots,4$), 
which are mixtures of the bino, the wino triplet and the higgsinos, 
which in turn are the superpartners of the U(1)$_Y$ and SU(2)$_L$ gauge bosons and Higgs doublets of \supersymmetry.

The primary motivation for the search for electroweakinos comes from its complementarity to the strong-production searches. 
In fact, electroweak production may be the dominant production mode for supersymmetric particles at the LHC if the squarks and gluinos are sufficiently heavy.
Another motivation is naturalness \cite{Barbieri:1987fn,deCarlos:1993yy}: 
Natural models of \supersymmetry suggest that the masses of the lightest charginos and neutralinos 
  fall into a range that is well accessible at the LHC.

All searches for \supersymmetry have produced null results so far, 
  as no significant excess beyond the event yields expected from Standard Model processes has been observed.
These null results can be translated into limits on the masses of supersymmetric particles in simplified models,
  which for strongly produced particles reach beyond \TeV{1},
  whereas in the case of electroweak production they are of the order of several hundreds of GeV. %
Recent results from the electroweak analyses carried out by the ATLAS collaboration include
  a search for exotic decays of the observed \GeV{125} Higgs boson into light neutralinos and possibly gravitinos.
This yields final states with photons and large missing transverse momentum (\met) %
  and is motivated from Gauge-Mediated Supersymmetry Breaking and Next-to-Minimal Supersymmetry extensions of the Standard Model \cite{ATLAS-CONF-2015-001}.
Then there is another analysis
  which does somewhat the opposite and looks 
  for the production of a chargino \charginon{1} and a neutralino \neutralinon{2} decaying 
  via a $W$ boson and the \GeV{125} Higgs boson \cite{PaperWh}. 
The latest result from the ATLAS collaboration on the search for electroweakinos
  is the electroweak summary paper \cite{EWKSummary}. %
This paper will be the focus of these proceedings. %

\section{THE SEARCH FOR ELECTROWEAKINOS}
The goal of the electroweak summary paper is to summarize and to extend the searches for electroweak \supersymmetry 
  with the ATLAS detector using the data taken during the first run (Run-1) of the LHC,
  corresponding to \ifb{20} of $pp$ collisions at a center-of-mass energy \eighttev.
It is not only a summary 
  but also includes five new analyses that have not been published before. 
These analyses look for two- or three-lepton final states and strive to extend the reach of earlier analyses 
  by lowering the thresholds on the transverse momenta (\pt) of the leptons, by exploiting initial-state radiation (ISR) or vector-boson fusion (VBF) event topologies, 
  or through the application of multi-variate analysis (MVA) techniques.
In addition to these five new analyses, 
  statistical combinations of the new and the existing searches are performed to extend the excluded mass ranges, 
  adding also new combinations and reinterpretations of existing searches.
Furthermore, the impact of the assumption for the mass of the intermediate slepton in simplified models on the exclusion reach is studied.
One particular focus of the new analyses is to improve the sensitivity of the searches
  for \supersymmetry scenarios with compressed mass spectra,
  where small mass differences between the particles in the \supersymmetry decay chains
  lead to low-energetic decay products.
Due to their low energy and momentum,
  these decay products may fail trigger or offline thresholds 
  and may thus not be reconstructed.
This deteriorates the signal acceptances 
  and reduces the sensitivity of the analyses.
\begin{figure}[t]
  \centerline{
    \incgraphics{width=3.2cm}{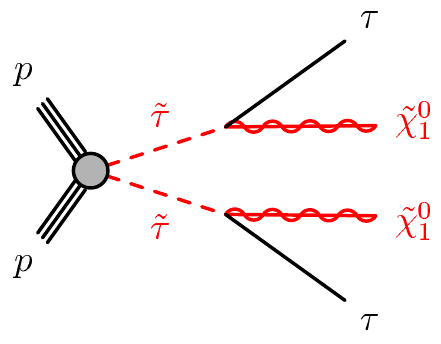}
    \incgraphics{width=3.2cm}{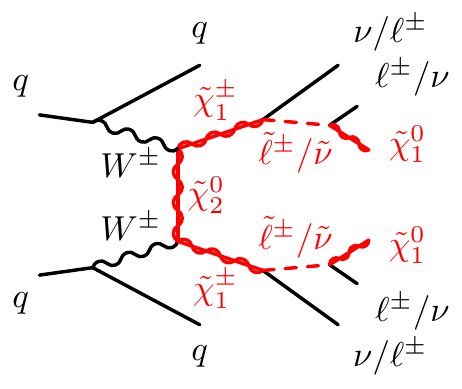}
    \incgraphics{width=3.2cm}{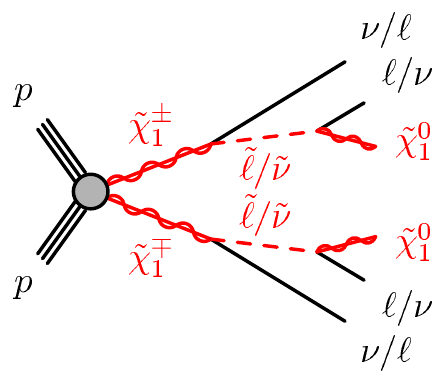}
    \incgraphics{width=3.2cm}{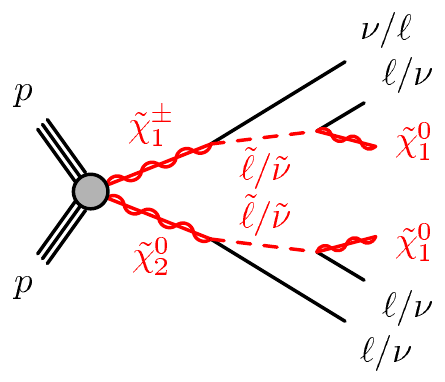}
    \incgraphics{width=3.2cm}{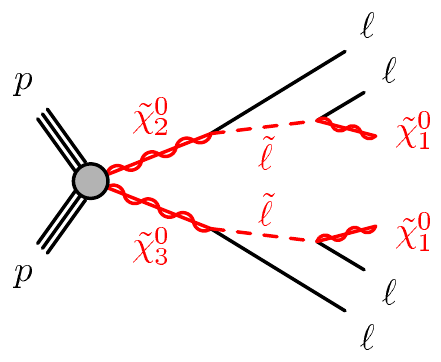}
  }
  \caption{
    Diagrams illustrating the production of supersymmetric particles and their decays modes 
    in the five simplified models that are employed in the interpretation of the results in the electroweak summary paper.
  }
  \label{fig:simple_models}
\end{figure}
The results of the searches described in the electroweak summary paper are interpreted in two classes of models.
The first class are simplified models, 
  where only one specific production mode and decay chain for the supersymmetric particles is considered, 
  and the branching ratios for the decays are assumed to be \percent{100}. 
Five simplified models are employed in the interpretation of the results in the electroweak summary paper 
  as shown in \Fig{simple_models} (from left to right):
  Production of stau pairs, 
  production of same-sign chargino pairs in a VBF topology, 
  chargino-pair production, 
  associated production of the lightest chargino (\charginon1) and the second-lightest neutralino (\neutralinon2), 
  and production of \neutralinon2 together with \neutralinon3. 
In all cases, the decays of the electroweak gauginos may proceed via all three slepton or sneutrino generations.
The second class of models are phenomenological models:
In the electroweak phenomenological Minimal Supersymmetric extension of the Standard Model (pMSSM),
  only the direct production of charginos and neutralinos is considered,
  which results in a small number of only four parameters. %
The two-parameter Non-Universal Higgs Masses model (NUHM2) is basically a constrained MSSM with two additional parameters 
  that allow to tune the Higgs masses. 
The third phenomenological model used in the paper
  is a scenario with Gauge-Mediated Supersymmetry Breaking (GMSB), 
  where the lightest supersymmetric particle (LSP) is the gravitino.
This is different from all other models that are considered, 
  where the LSP is always the lightest neutralino \neutralinon{1}. 
For the GMSB scenario, electroweak production dominates for large values of the parameter $\Lambda$,
  the \supersymmetry-breaking mass scale felt by the low-energy sector. %
In all models discussed in these proceedings, $R$-parity is assumed to be conserved.

Two of the five new analyses are independent from the others 
  as their selections have little overlap with the other analyses
  and their results are interpreted in models specific to these analyses.
These two analyses are the two-tau analysis using an MVA technique
  and the search for two same-sign (SS) leptons in a vector-boson fusion topology.
They will therefore be discussed, including their results and interpretations,
  separately from the others.

\subsection{2$\tau$ (MVA) Analysis} %
\begin{figure}[tbh]
  \centerline{
    \includegraphics[width=10cm]{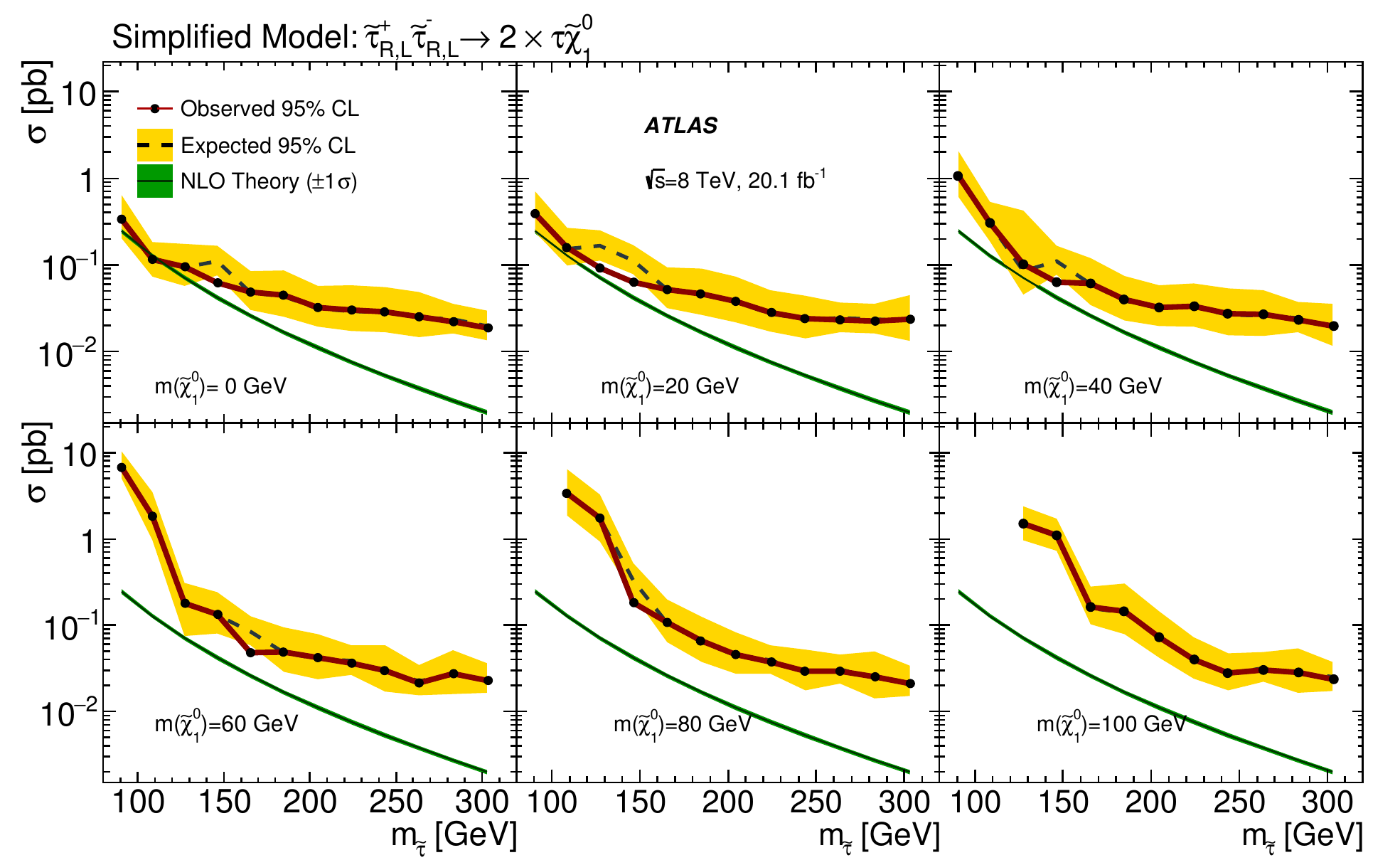} %
  }
  \caption{
    For six different masses of the LSP,
    these plots show the \percent{95} CL exclusion limits 
    on the cross sections for combined pair production of left- and right-handed staus
    as a function of the stau mass
    \cite{EWKSummary}.
  }
  \label{fig:result_twotau}
\end{figure}
The \analysis{two-tau (MVA)} analysis is an update of an earlier analysis,
  which targets the direct production of charginos, neutralinos and staus 
  in final states with at least two hadronically decaying tau leptons and missing transverse momentum \cite{Paper2tau}.
In contrast to the simpler, cut-based analysis,
  the updated version makes use of a boosted-decision tree (BDT) to improve the sensitivity,
  and the results are %
  interpreted in a simplified model with direct-stau production,
  where the cut-based analysis had practically no sensitivity. 
In the two-tau (MVA) analysis, events with exactly %
  two taus with opposite charge are selected.
Events that contain a $b$-tagged jet or where the two taus have an invariant mass 
  that is compatible with the assumption that the taus are coming from a $Z$-boson decay
  are rejected.
A boosted decision tree is trained on twelve input variables,
  which are based on kinematic properties of the two taus and \met
  and have good discriminatory power.
A cut on the output value of the BDT is then used to define one signal region (SR).
Good agreement of the distribution of the BDT output variable in data
  and its expected distribution from the Standard Model background prediction is found 
  prior to the signal-region cut,
and no excess is observed in the signal region.
This allows to interpret the analysis in terms of limits on the production cross section for stau pairs
  as a function of the mass of the stau and the LSP,
  as shown in \Fig{result_twotau}.
The best limit is found for a stau mass around \GeV{110} and a massless LSP.
\subsection{2 SS \lepton (VBF) Analysis}
The second analysis 
  from the summary paper  %
  to be presented here 
  looks at final states with two light leptons with the same charge.
It specifically targets a scenario where supersymmetric particles are produced via vector-boson fusion (VBF)
  as shown in the second diagram in \Fig{simple_models}.
This reduces the production cross section considerably %
  but on the other hand makes it easier to separate signal and background 
  by requiring the presence of the two additional VBF jets.
The jets also often cause the chargino to be boosted,
  yielding energetic decay products even in compressed spectra.
In addition to two light leptons with the same charge, 
  events selected in this analysis are required to have two jets and missing transverse momentum above \GeV{120}
  in order to be in the plateau of the \met trigger that is used in this analysis.
(This is a unique feature of this analysis. 
In contrast, all of the other four analyses use combinations of single, double, and triple lepton triggers.)
One cut-based signal region is defined,
  exploiting the VBF topology 
  by requiring that the two jets be well-separated and in opposite hemispheres of the ATLAS detector
  and have a large invariant mass.
Additional cuts suppress the remaining Standard Model backgrounds, mainly diboson and top quark production. %
No excess is observed in the SR, 
  thus limits are set on the VBF \ConeSS production cross section.
The limits obtained from the 2012 dataset remain above the theoretical predictions by at least a factor three, %
  \ie this analysis is not yet sensitive to VBF \ConeSS production.
Exclusion plots for two different assumptions 
  on the mass of the lighter chargino \charginon{1}
  and as function of the mass splitting $\Delta m(\charginon{1}, \neutralinon{1})$ 
  can be found in the paper \cite{EWKSummary}.
CMS recently made public a search
  that is able to set limits on the electroweak production of supersymmetric particles in a VBF scenario \cite{CMS-PAS-SUS-14-005}.
The main differences are that the CMS search 
  does not only consider same-sign chargino production but combines several production modes,
  assumes larger mass splittings between the chargino and the neutralino,
  and decays to happen via staus only.

\subsection{Compressed Spectra With Two- and Three-Lepton Final States}
In the following part, the remaining three of the five new analyses %
  are discussed,
  before then coming the a joint presentation of their results and interpretations.
These three analyses look at final states with two or three light leptons 
  and aim to extend the reach of earlier searches for the production of electroweakinos \cite{Paper2l,Paper3l}
  for compressed \supersymmetry scenarios.

The \analysis{2 OS \lepton (ISR)} analysis %
  extends the earlier search for \supersymmetry in final states with two leptons \cite{Paper2l}
  to small mass splittings $\Delta m(\charginon{1}, \neutralinon{1})$ 
    between the lightest chargino \charginon{1} and neutralino \neutralinon{1} %
  by exploiting initial-state radiation jets.
The ISR jet boosts the leptons from the supersymmetric decay chain,
  which otherwise may have too low momentum to pass the trigger or reconstruction thresholds. 
Events which have exactly two light leptons with opposite charge (OS leptons)
  and an ISR jet with high transverse momentum are selected,
  excluding events that contain $b$-tagged or forward jets,
  or in which the invariant mass of the two light leptons is close to the $Z$-boson mass.
Two signal regions are defined based on ``super-razor variables'' \cite{PhysRevD.89.055020} 
  with good discriminatory power in compressed spectra
  and the ratio $R_2$ of the missing transverse momentum and the sum of missing transverse momentum and the transverse momenta of the leptons. %
ISR jets are also used in the \analysis{3\lepton (ISR)} analysis,
  which extends the corresponding earlier search \cite{Paper3l}
  to small mass splittings $\Delta m(\neutralinon{2}, \neutralinon{1})$
    between the second-lightest neutralino \neutralinon{2} (or lightest chargino \charginon1) and the lightest neutralino \neutralinon{1}.
Moreover, three-lepton triggers are now included 
  which allow to go lower in lepton \pt (\analysis{soft leptons}).
Events must have exactly three light leptons, including one pair with same flavor but opposite charge (SFOS).
After a veto on events with $b$-tagged jets or where a SFOS lepton pair comes from an $\Upsilon$ meson decay,
  four signal regions are defined 
  which either veto or require a jet with large \pt 
  and differ in the allowed window for the value of %
    the minimum mass of the SFOS pairs.
Finally, the \analysis{2 SS \lepton (MVA)} analysis complements the search for three-lepton final states 
  in case one of the three leptons is missed,
  selecting events with exactly two light leptons with the same charge sign.
This analysis makes use of eight boosted decision trees
  which are trained independently to define the same number of signal regions,
  optimized for four different mass-splitting scenarios $\Delta m(\neutralinon{2}, \neutralinon{1})$,
  each with and without the presence of ISR jets.
The output of the BDT is also used to define validation regions
  that demonstrate that the Standard Model backgrounds are well understood.
All three analyses observe good agreement between the event counts in data and their Standard Model predictions
  and no significant excess in any of the signal regions.
These results are interpreted in terms of exclusion limits,
  combining analyses where they have comparable sensitivity.

\section{INTERPRETATIONS}

\begin{figure}[tb]
  \centerline{
    \incgraphics{width=5.75cm,clip,trim=0 20 0 0}{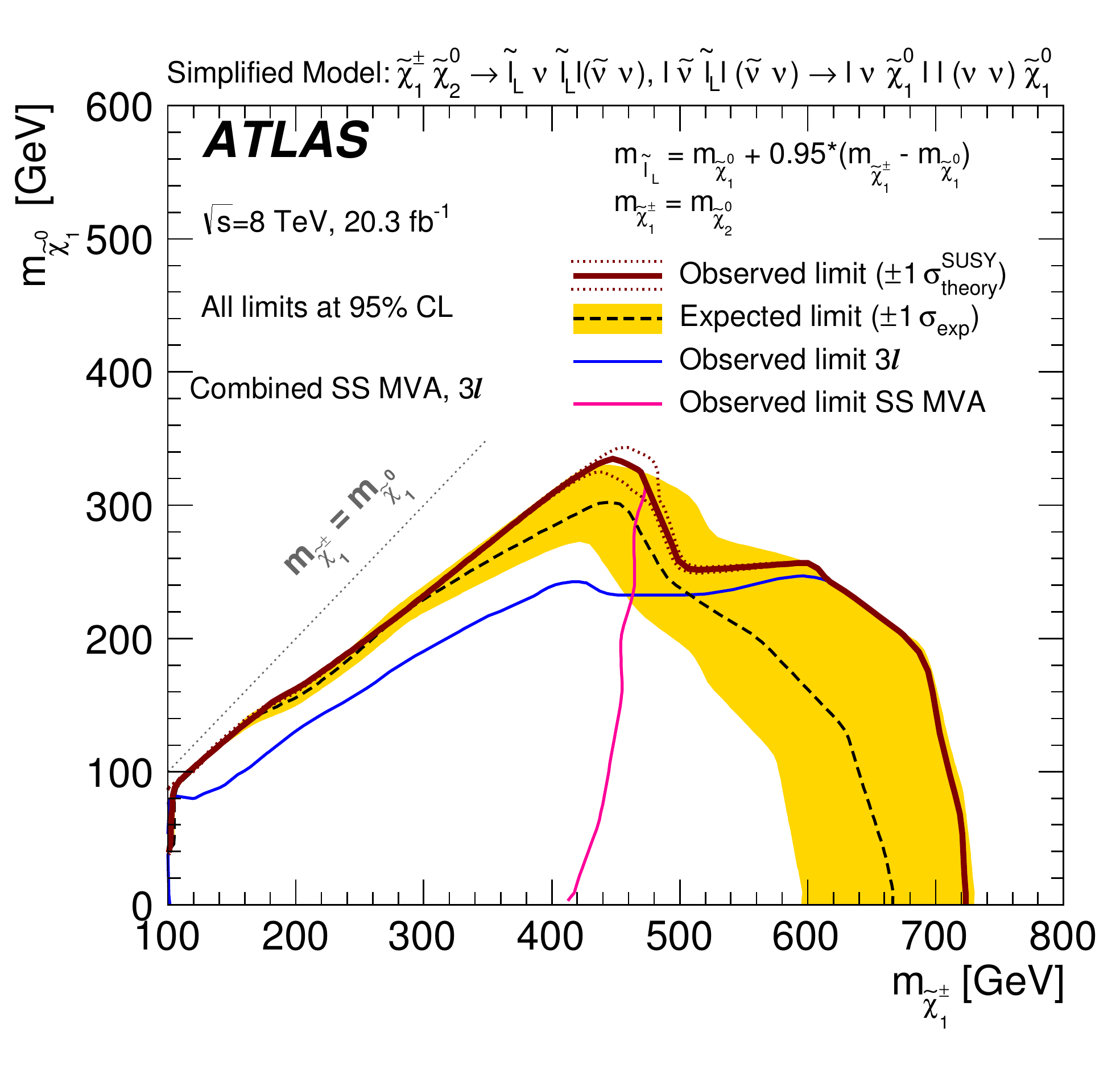} %
    \incgraphics{width=5.8cm,clip,trim=0 20 0 0}{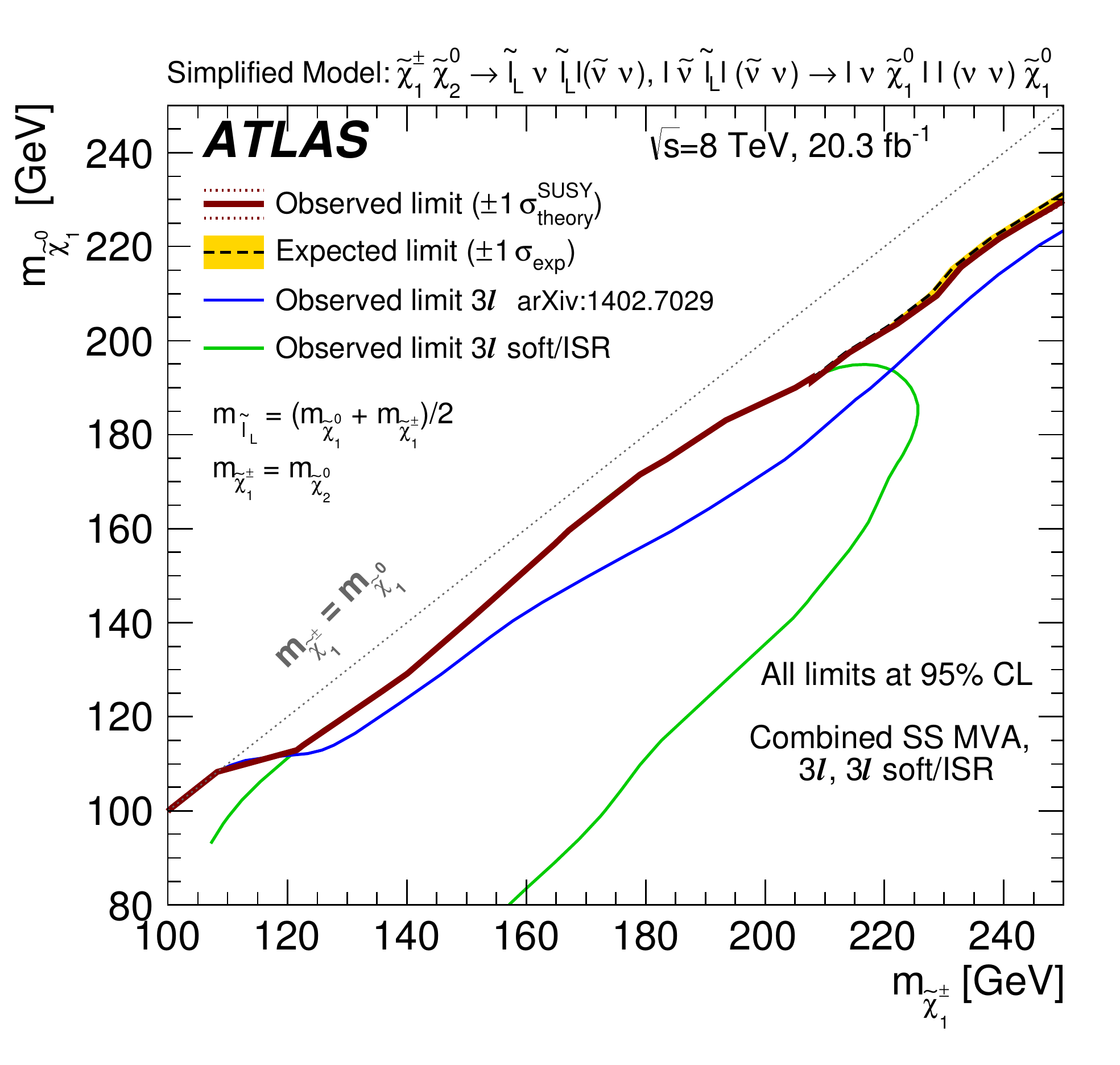} %
    \incgraphics{width=5.25cm}{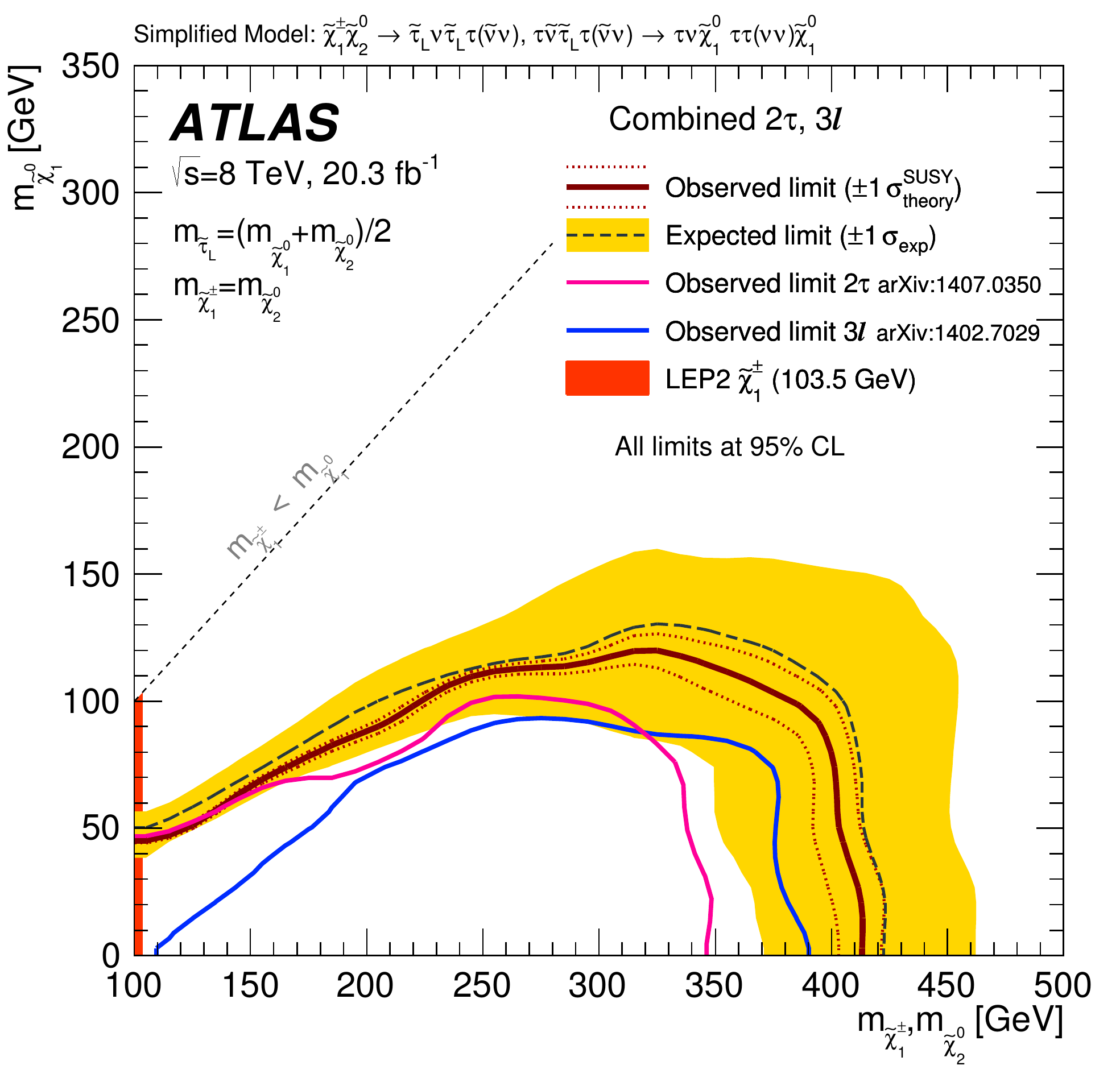} \\ %
  }
  \caption{
    The \percent{95} CL exclusion limits on \ConeNtwo production from the combinations of several analyses \cite{EWKSummary}.
    Left: $\slepton_L$-mediated decays with sleptons close in mass to the \neutralinon{2},
    middle: zoom of the compressed region, $\slepton_L$-mediated decays with sleptons masses halfway between \neutralinon{2} and \neutralinon{1},
    right: $\stau_L$-mediated decays with stau mass halfway between \neutralinon{2} and \neutralinon{1}.
  }
  \label{fig:interpretations_c1n2}
\end{figure}

The exclusion limits on the mass parameters $m(\neutralinon{1})$ and $m(\charginon{1}) = m(\neutralinon2)$ 
  for the simplified model with \ConeNtwo production 
  are shown in the plots in \Fig{interpretations_c1n2}.
Both \charginon{1} and \neutralinon{2} are assumed to be pure wino,
  whereas the \neutralinon{1} is pure bino.
The resulting observed limits from combinations of both the new analyses 
  and the ones published earlier as indicated in the plots,
  are given by the thick red lines
  and compared against the earlier exclusion limits,
  drawn as thinner lines.
The left plot in the Figure shows the complementarity of the new 2 SS \lepton (MVA) analysis 
  and the earlier 3\lepton analysis.
In the middle, a zoom-in of the compressed region close to the diagonal is shown,
  which highlights the improvement of the limit 
  in this difficult region
  that is obtained from the new 3\lepton (ISR / soft-leptons) analysis.
Note that the assumption on the mass of the sleptons here is different from the one in the left plot.
  The 2 SS \lepton (MVA) analysis is also included in the combination
  but has no sensitivity in this region by itself 
  and thus no exclusion line is shown. %
The right-hand plot of \Fig{interpretations_c1n2}
  shows a new combination of the existing 2$\tau$ and 3\lepton analyses in a scenario 
  where the gaugino decays are mediated via staus only.

\begin{figure}[t]
  \centerline{
    \incgraphics{width=6.05cm,clip,trim=0 20 0 0}{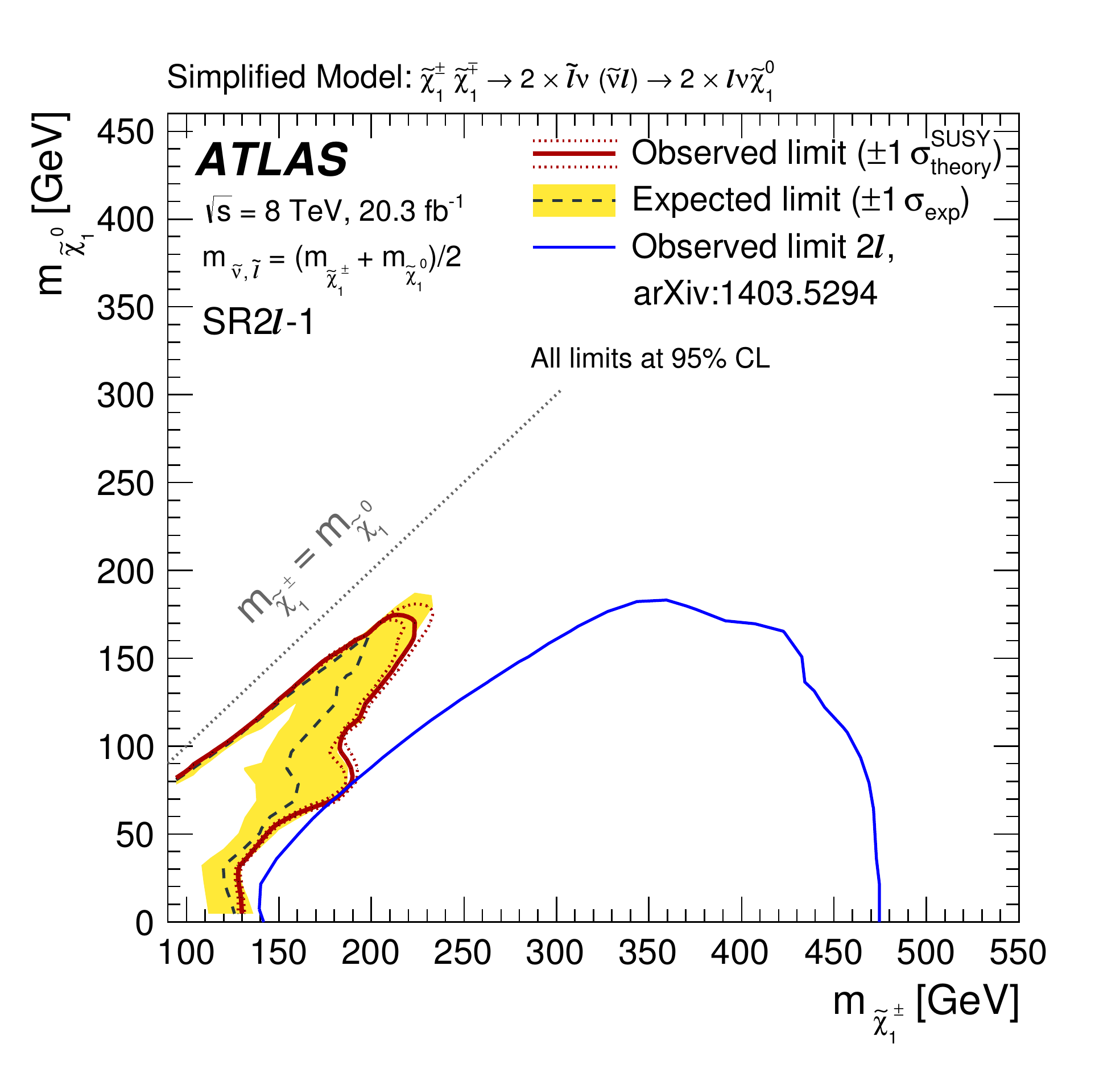} %
    \incgraphics{width=5.95cm}{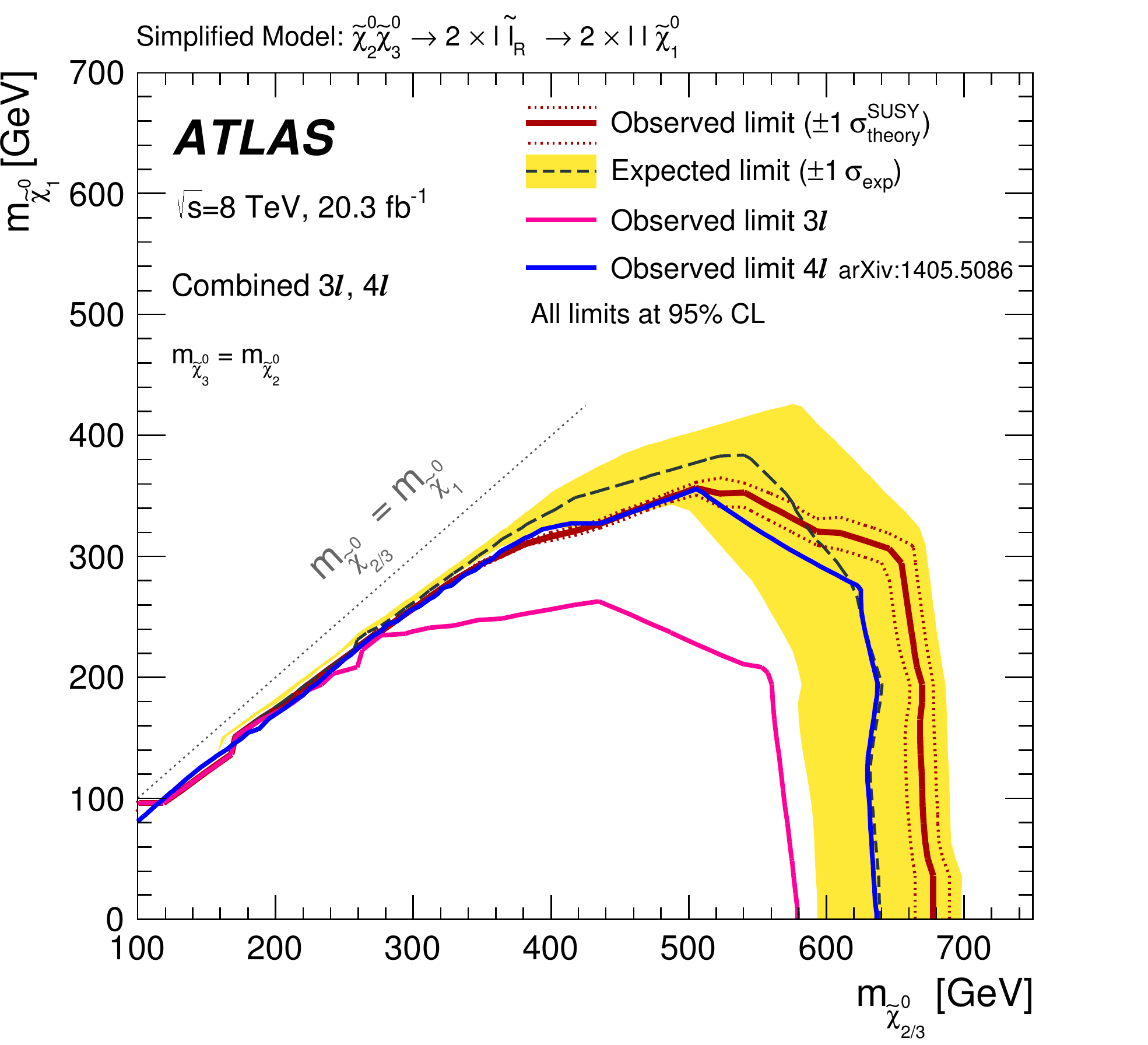} %
    \incgraphics{width=5.7cm,clip,trim=0 20 0 0}{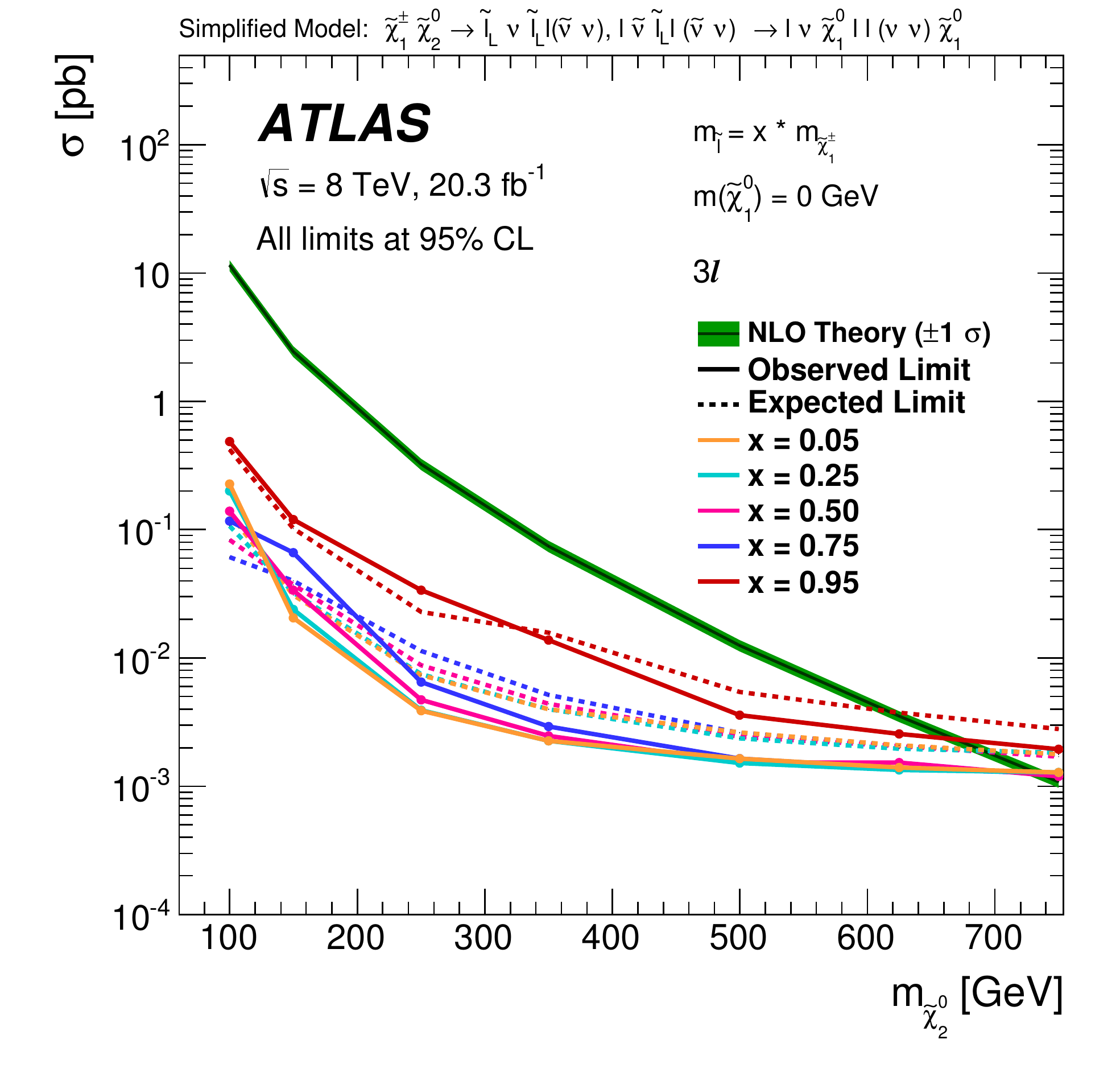} %
  }
  \caption{
    The \percent{95} CL exclusion limits 
      on \ConeOS production (left) 
      and \NtwoNthree production (middle) 
      with slepton-mediated decays.
    The right-hand plot shows the impact of the intermediate slepton mass on the exclusion limit
    \cite{EWKSummary}.
  }
  \label{fig:interpretations2}
\end{figure}

The left plot in \Fig{interpretations2} demonstrates
  the complementarity of the new 2 OS \lepton (ISR) analysis 
    and the earlier 2\lepton analysis
  in the exclusion limits for the \ConeOS simplified model
  with $\slepton_L$-mediated decays,
  where the new analysis fills %
    the gap
    between the existing exclusion contour and the diagonal kinematic boundary.
Again, the \charginon{1} is a pure wino and the \neutralinon1 a pure bino in this simplified model.
The plot in the middle shows a new combination 
  of the existing 3- and 4-lepton analyses
  in the simplified model with \NtwoNthree production 
  and decays mediated via $\slepton_R$.
Here, the \neutralinon{2} and \neutralinon{3} are assumed to be pure higgsino and mass-degenerate.
This combination improves the earlier limits on the mass of the initial supersymmetric particles
  from the 4-lepton analysis by about \GeV{30}.
For all three simplified models, \ConeOS, \ConeNtwo, and \NtwoNthree,
  the impact of the assumption for the intermediate slepton mass
  on the exclusion reach has been checked for a massless LSP
  by varying the slepton mass between $5$ and \percent{95} of the mass of the decaying gaugino.
The impact is found to be small,
  as can be seen in the right plot of \Fig{interpretations2} for the case of \ConeNtwo.

\begin{figure}[h]
  \centerline{
    \incgraphics{width=6.0cm}{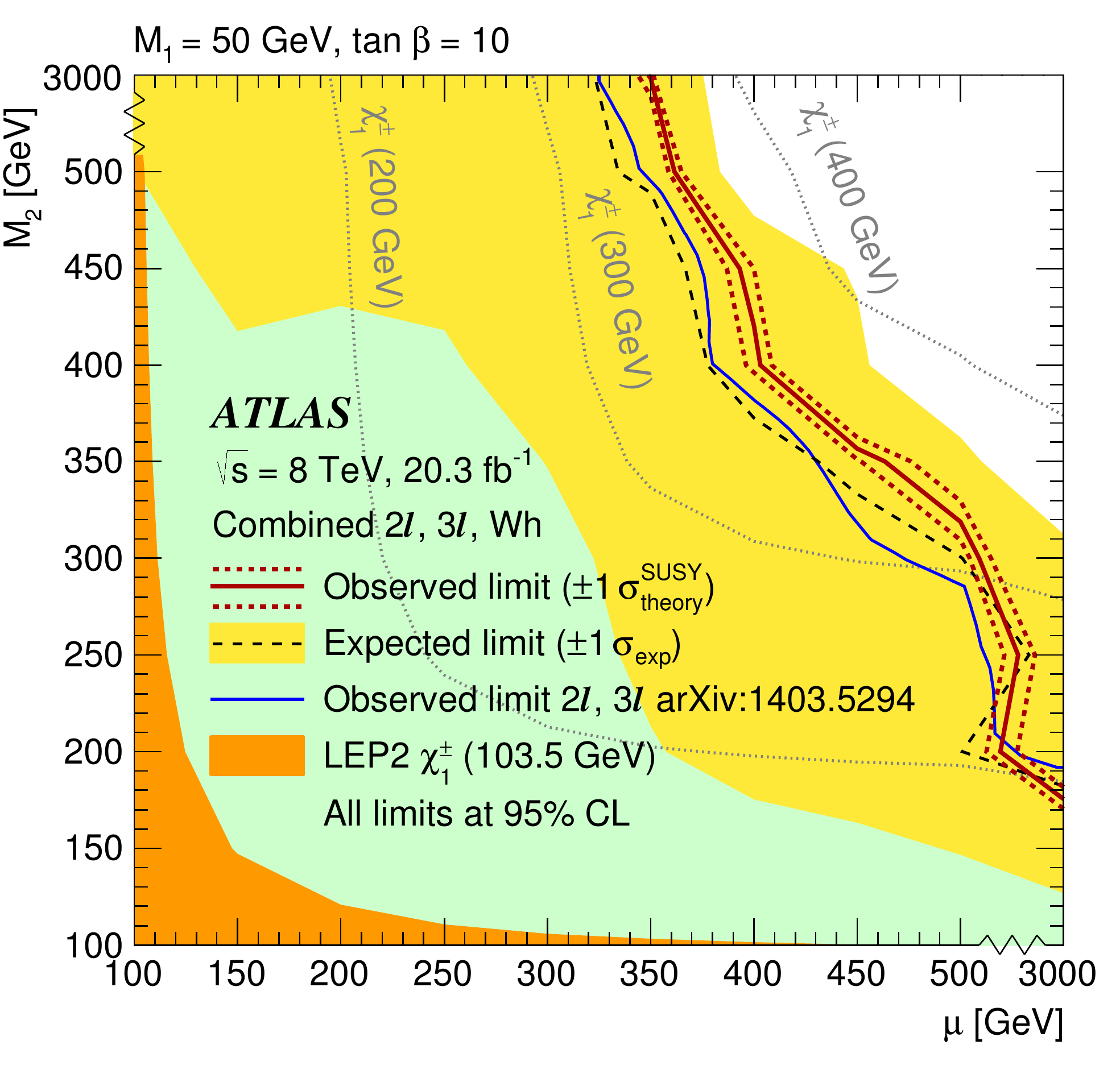} %
    \incgraphics{width=5.9cm}{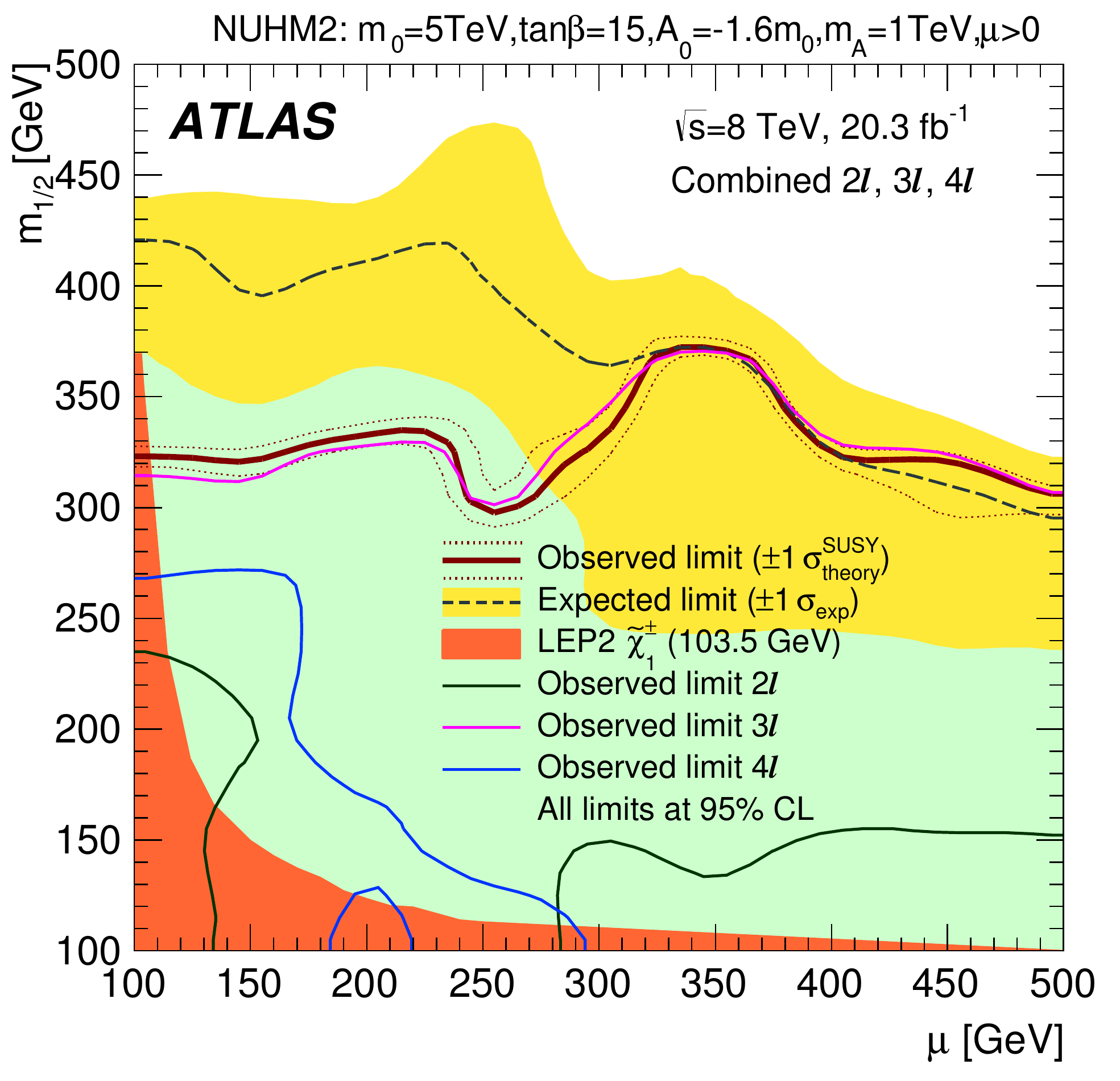} %
    \incgraphics{width=6.33cm}{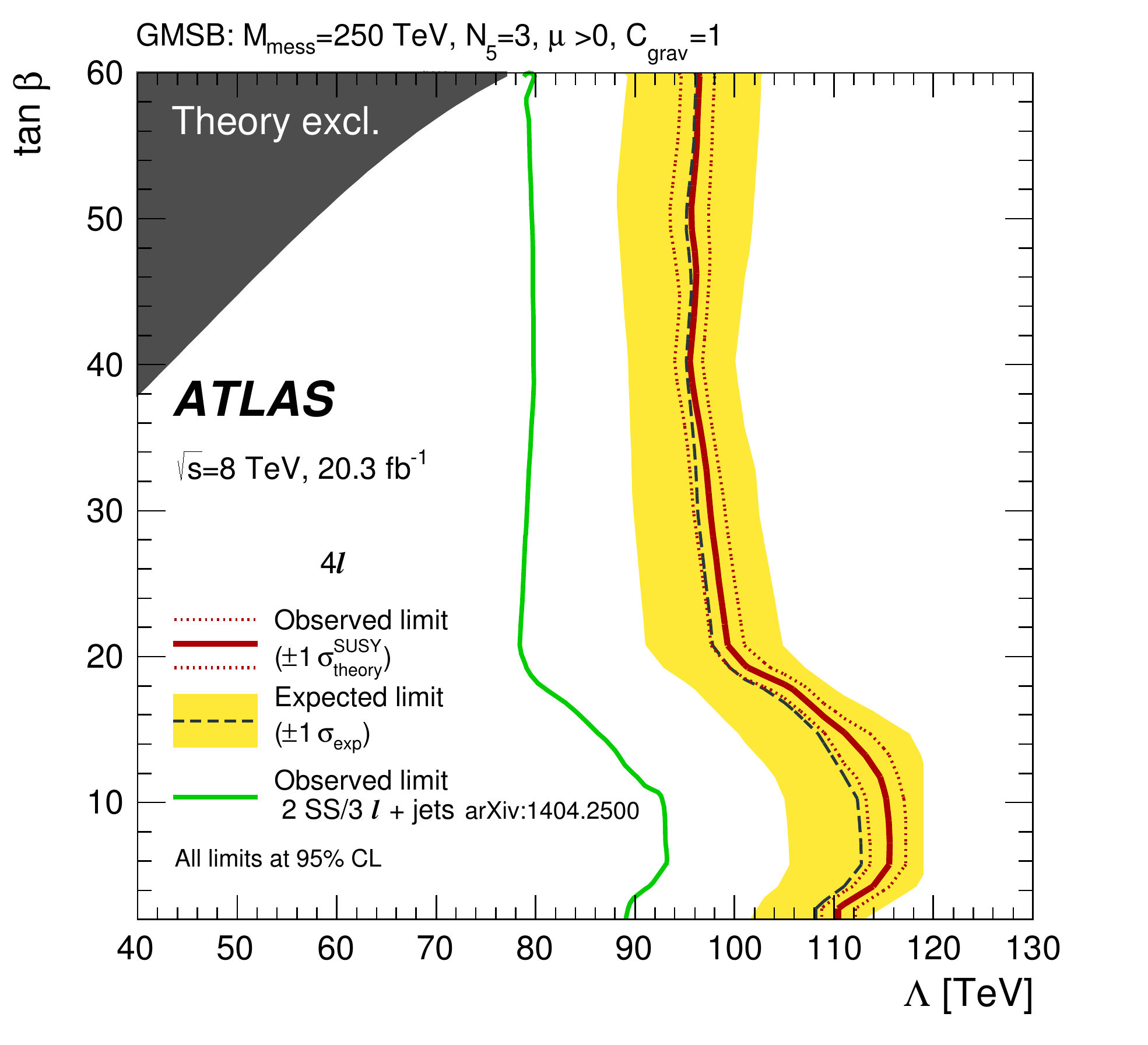} %
  }
  \caption{
    \percent{95} CL exclusion limits from the searches for electroweakinos in phenomenological models:
    pMSSM (left),
    NUHM2 (middle), 
    and GMSB (right) 
    \cite{EWKSummary}.
  }
  \label{fig:interpretations_pheno}
\end{figure}

The results of the electroweak analyses have also been interpreted in terms of exclusion limits
  on parameters of phenomenological models.
These limits are shown in \Fig{interpretations_pheno}.
The \percent{95} CL exclusion limit
  on the pMSSM parameters $\mu$ and $M_2$ 
  is obtained by combining the $Wh$ analysis \cite{PaperWh}
  with the results from the 2\lepton and 3\lepton analyses \cite{Paper2l,Paper3l}
  and is shown in the left plot.
The limits on the NUHM2 parameters $\mu$ and $\moh$ 
  come from a new combination of the existing 2\lepton, 3\lepton, and 4\lepton analyses
  and are shown in the middle plot in \Fig{interpretations_pheno}.
It can be seen that the 3\lepton analysis drives the exclusion limit in this scenario.
The right plot shows that
  a new reinterpretation of the 4\lepton analysis \cite{Paper4l}
  yields an improvement of 15 to \TeV{20}
  with respect to an earlier combination \cite{PaperStrong2SS3L}. %
The electroweak summary paper also has two plots
  which compare all earlier and new exclusion contours 
  in the mass-parameter planes of the respective simplified models,
  providing separate comparisons for slepton- and $W$ / $Z$ / $h$-boson mediated decays.
For the latter decay modes,
  the sensitivity of the new analyses is small,
  thus no combination has been attempted.

\section{CONCLUSION}
In conclusion, now that the searches for \supersymmetry in the data from the first run of the LHC have been wrapped up,
  no strong signs for physics beyond the Standard Model have emerged from the Run-1 data.
Still, a large number of \supersymmetry analyses have been made public that constrain the parameter space of supersymmetric models.
In these proceedings, the new electroweak summary paper has been discussed,
  which contains the final ATLAS limits  
  on the electroweak production of supersymmetric particles
  at a \com energy of \eighttev.
It should be stressed that this paper is not only a summary, but also includes completely new analyses,
  explores several new analysis techniques,
  and includes new combinations.

The second run of the LHC has started,
  and a dataset of around \ifb{4} of integrated luminosity from proton-proton collisions 
  at the increased \com energy of \tttev has been collected in 2015.
As the cross sections for the production of heavy particles grows 
  stronger than linearly with the \com energy,
the higher the relevant mass ranges,
  the stronger the benefit from the increased \com energy for the expected reach of a search.
The first results with the new data will thus come from the strong-production searches,
  which benefit a lot more from the higher \com energy
  as the larger cross sections for strong production allow them to go higher in mass 
  than the electroweak searches.
Electroweak searches in general need more data 
  to improve upon with the existing Run-1 results.
However, they will be able to build upon lots of experience gained during Run-1,
  and on the long term will also profit from the higher integrated luminosity 
  to be collected in the three years of Run-2,
  which is expected to exceed the integrated luminosity from Run-1 by a factor around four.

\bibliographystyle{aipnum-cp}%
\bibliography{proceedings_arxiv}%

\end{document}

%% file: befehle_arxiv.tex
\newcommand{\kommentar}[1]{}

\newcommand{\inmath}[1]{\ensuremath{#1}\xspace}

\makeatletter
\DeclareRobustCommand*{\trigger}[1]{%
  \begingroup\@activeus\scantokens{#1 }\endgroup}
\begingroup\lccode`\~=`\_\relax
   \lowercase{\endgroup\def\@activeus{\catcode`\_=\active \let~\_}}
\makeatother

\newcommand{\met}{$E_{\textnormal{T}}^{\textnormal{miss}}$\xspace}

\newcommand{\pt   }{\inmath{p_\text{T}}} %

\newcommand{\GeV}[1]{{\unit[#1]{GeV}}}
\newcommand{\TeV}[1]{{\unit[#1]{TeV}}}

\let\etaorig\eta%
\renewcommand{\eta}{\ensuremath{\etaorig}}

\newcommand{\eighttev}{\inmath{\sqrt{s} = \TeV{8}}}
\newcommand{\thirteentev}{\inmath{\sqrt{s} = \TeV{13}}}
\newcommand{\tttev}{\thirteentev}

\newcommand{\onehalf      }{\inmath{{\nicefrac12}}}

\newcommand{\neutralinon  }[1]{\inmath{\widetilde{\chi}^0_{#1}}}

\newcommand{\charginon    }[1]{\inmath{\widetilde{\chi}^\pm_{#1}}}

\newcommand{\charginomp   }   {\inmath{\widetilde{\chi}^\mp}}

\newcommand{\stau         }{\inmath{\widetilde{\tau}}}

\newcommand{\lepton       }{\inmath{\ell}}
\newcommand{\slepton      }{\inmath{\widetilde{\lepton}}}

\newcommand{\ifb}[1]{\inmath{\unit[#1]{fb^{-1}}}} %

\newcommand{\moh          }{\inmath{{m_\onehalf}}}

\newcommand{\bit}{\begin{itemize}}
\newcommand{\eit}{\end{itemize}}
\newcommand{\bfr}[1]{\begin{frame}[squeeze] \frametitle{#1}}

\newcommand{\ie}{i.\,e.\@\xspace}

\newcommand{\percent}[1]{{#1}\,\%}